\documentclass[a4,10pt,onecolumn]{article}
\usepackage{graphicx} 
\usepackage[margin=0.8in]{geometry} 
\usepackage{amsmath,amsthm,amssymb}
\usepackage{float}
\usepackage{xcolor}
\usepackage[utf8]{inputenc}
\usepackage{graphicx}
\usepackage[backend=bibtex,style=nature]{biblatex}
\usepackage{lineno}
\usepackage{authblk}

\def\l#1{\label{eq:#1}}
\def\r#1{(\ref{eq:#1})}

\def\_#1{{\bf #1\mit}}

\graphicspath{{Figures_temp/}{Figures_final/}}
\usepackage[font=footnotesize]{caption}

\title{Giant bias-free nonreciprocity for unpolarized light via synthetic motion}

\date{}
\author[1]{L.M. Máñez-Espina$^*$}
\author[2]{B. Amrahi$^*$}
\author[3]{I. Faniayeu}
\author[3]{R. Cichelero}
\author[3]{A. Dmitriev}
\author[1]{A. Díaz-Rubio}
\author[2]{V.S. Asadchy}
\affil[1]{Nanophotonics Technology Center, Universitat Politècnica de València, València 46022,  Spain.}
\affil[2]{Department of Electronics and Nanoengineering, Aalto University, Espoo, 02150, Finland.}
\affil[3]{Department of Physics, University of Gothenburg, Gothenburg, 41296, Sweden}

\begin{document}

\maketitle

$^*$These authors have contributed equally

\begin{abstract}
Reciprocity breaking at optical frequencies typically relies on bulky magnets, dynamic modulation, or nonlinearities, all of which hinder chip-scale integration and the handling of unpolarised light. We introduce a fully passive, subwavelength metasurface that achieves polarisation-insensitive one-way transparency by combining self-magnetised ferrite nanodisks in a vortex state with symmetry-protected quasi–bound states in the continuum. The metasurface exhibits a pure synthetic moving-medium response at optical frequencies, yielding giant nonreciprocal directional dichroism.
We report near-unity values for both the transmittance contrast and the emissivity-to-absorptivity ratio with experimentally widely available ferrite materials, all under unpolarised illumination and without external bias. Using temporal coupled-mode theory, we identify the design conditions necessary to maximise directional dichroism: critical coupling, Huygens-type resonance overlap, and strong inter-mode coupling.
Furthermore, we propose a deterministic, stamp-assisted protocol for imprinting arbitrary, uniform or patterned vortex configurations across large arrays of nanodisk meta-atoms, enabling scalable fabrication.
This work establishes a practical route toward compact nonreciprocal photonics with applications in photonic gyrators, nonreciprocal wavefront engineering, and nonreciprocal solar cell technologies.
\end{abstract}


Reciprocity is a fundamental concept in science used in various fields, from quantum mechanics~\cite{Dek2012} to game theory~\cite{Falk2006}. In electrodynamics, a system is said to be Lorentz reciprocal when the measured fields created by a source at an observation point are the same as in the case where the locations of the source and observation point are interchanged~\cite{Potton2004,Caloz2018,Asadchy2020}. Breaking Lorentz reciprocity in a system is crucial for many applications~\cite{Tsakmakidis2017,Born_Wolf_2019,Adam2002}. Nonreciprocal elements, such as optical isolators, provide a simple solution to prevent electromagnetic energy from back-feeding laser cavities or to isolate communication channels. In the last decade, nonreciprocal optical elements have also regained importance due to their pivotal role in advancing thermal photonics~\cite{Yang2024}. Indeed, emissivity and absorptivity can differ in nonreciprocal systems, as Kirchhoff's law of radiation stems from the Lorentz reciprocity~\cite{Park2021,Guo2022}. Therefore, nonreciprocal optical systems were proposed to improve energy harvesting in solar cells~\cite{Park2022,Green2012}, thermal management technologies~\cite{Khandekar2020,Fan2022}, thermophotovoltaics~\cite{JafariGhalekohneh2022,Park2022thermo}, and other thermal photonic technologies.

The conventional approach to achieving nonreciprocity in optical systems relies on the use of magneto-optical (MO) materials~\cite{Asadchy2020}. Although bulk natural materials exhibit only weak MO effects, these can be significantly enhanced in nanophotonics via optical resonances~\cite{Armelles2013,Chin2013,MacCaferri2015,Barsukova2017,Chernyak2020,Abujetas2021,Park2021,Xia2022,Mez-Espina2024,Ruan2025}. However, these material systems are inherently polarisation-dependent, providing the desired nonreciprocal response only for one of the two orthogonal polarisations. Since most optical applications involve unpolarised light, the contrast between transmittances in the opposite directions and the contrast between emissivity and absorptivity in such structures are fundamentally limited to 50\%. Furthermore, the operation of these designs requires an external magnetic field, complicating their implementation and rendering them unsuitable for many practical scenarios.
Recently, alternative approaches to breaking electromagnetic reciprocity have been proposed, based on time-modulated structures~\cite{Fan2012,Sounas2017,Shaltout2015} and nonlinear materials ~\cite{Mahmoud2015,Lawrence2018,tripathi_nanoscale_2024,Cotrufo2024}. The former remains highly challenging at optical frequencies due to the requirement of complex dynamic biasing, and the latter is fundamentally constrained by dynamic reciprocity~\cite{Shi2015}.

Here, we demonstrate polarisation-insensitive optical isolation with a theoretical limit of 100\% based on a subwavelength metasurface composed of self-magnetised ferrite nanodisks. All nanodisks exhibit a remanent magnetic vortex state with identical circularity, rendering the structure entirely passive and eliminating the need for external magnetic bias.
We propose a straighforward experimental methodology for realising such vortex-state metasurfaces using conventional ferrite materials and validate it through rigorous micromagnetic simulations.
By leveraging symmetry-protected quasi-bound states in the continuum (quasi-BICs~\cite{Hsu2013,Hsu2016}) and temporal coupled-mode theory (TCMT), we achieve a contrast in light transmittance between opposite propagation directions of up to 71\%, and a light absorptivity/emissivity ratio of up to 70\%, under unpolarised light excitation and using realistic material parameters.
Owing to its spatiotemporal symmetry, the metasurface exhibits a giant nonreciprocal directional dichroism (NDD)~\cite{BARANOVA1977,Barron1984,Rikken1997,Rikken2000,Barron_2004,Lehmann2019,Park2022non,yokosuk2020nonreciprocal} (also known as the synthetic moving medium effect~\cite{Tetryakov1998,Mazor2014,Mirmoosa2014,Radi2014,Vehmas_2014,Radi20166,Huidobro2019,Mazor2020,Radi2020,Huidobro2021})—an exotic phenomenon in condensed matter physics that arises only in systems breaking both spatial inversion and time-reversal symmetries. While the NDD has recently been observed at microwave frequencies~\cite{Kodama2024,Li2024}, it remains negligibly small in the optical regime, limiting practical applications. In contrast, our vortex-based metasurfaces exhibit record-high NDD approaching 0.78.

\section*{Results}

\begin{figure}[tb]
    \centering
    \includegraphics[width=0.9\linewidth]{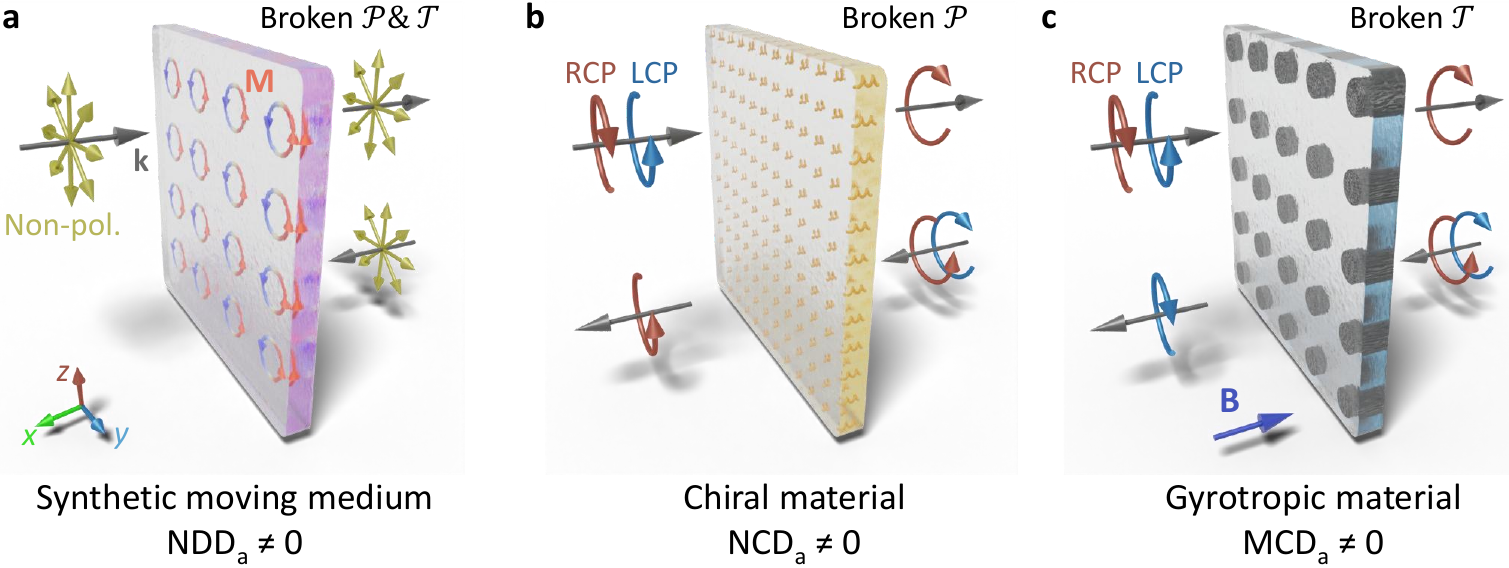}
    \caption{\textbf{ 
    Different dichroism effects and their physical origins.} 
\textbf{a,} polarisation-insensitive nonreciprocal directional dichroism, characterised by differing transmissivity for light incident from opposite directions. This effect arises from the synthetic moving medium mechanism and requires the breaking of both parity and time-reversal symmetries. In this work, the synthetic moving medium effect is realised using vortex magnetisation $\_M$, as indicated by the circular arrows within the material slab. 
\textbf{b,} Natural circular dichroism, which occurs in reciprocal materials with broken parity symmetry, resulting in direction-independent preferential absorption of light with a specific polarisation handedness. The illustration shows a chiral material slab embedded with helical inclusions.  
\textbf{c,} Magnetic circular dichroism, observed in nonreciprocal materials with broken time-reversal symmetry. As in \textbf{b}, light of a particular handedness predominantly transmits through the slab; however, the handedness that is transmitted depends on the direction of incidence. An example is illustrated with a Faraday material slab composed of an array of ferromagnetic cylinders magnetised by an external magnetic field $\_B$ oriented out-of-plane.
}
    \label{fig:Panel1}
\end{figure}

\subsection*{Directional and circular dichroisms and their origins}


NDD is defined as the difference in optical absorption or transmission when the same light beam travels through a medium in opposite directions.  NDD is typically defined for unpolarised light as    $\mathrm{NDD_i}=(T_{xx}^++T_{yy}^+ -T_{xx}^- -T_{yy}^-)/2$~\cite{Park2022non,yokosuk2020nonreciprocal}. Here, $T_{ij}$ are the transmittances of the structure that can be calculated from the complex transmission coefficients as $T_{ij}=|t_{ij}|^2$. The superscripts `$+$' and `$-$' define the opposite propagation directions along the $z$-axis, while the subscript `${\rm i}$’ in $\mathrm{NDD_{i}}$ indicates that the quantity is defined as the difference between light \textit{intensities} rather than complex field amplitudes. 
In the extreme case when NDD is close to unity, the material is highly transparent when measured with light in the $\textbf{k}=k \hat{z}$ direction but nearly opaque in the $\textbf{k}=-k \hat{z}$ direction ($k$ is the wavenumber), which is referred to as the effect of one-way transparency and depicted in Fig.~\ref{fig:Panel1}a. 
To obtain NDD, it is necessary to break both time-reversal symmetry, represented by the symmetry operator $\mathcal{T}$, and parity, $\mathcal{P}$. 
In the condensed-matter literature, NDD encompasses multiple effects, differentiated by the biasing mechanisms of the medium~\cite{Park2022non,Cheong2018}, including magnetochirality, transverse magnetochirality, and toroidal magnetic dichroism. Within the framework of bianisotropic material classification~\cite{575ea75d6592451b92c66b9cee72e6ad,463d18444148452ea4bc91b24b1d7c65}
\begin{equation}
    \_D = \overline{\overline{\varepsilon}} \cdot  \_E + \overline{\overline{\xi}} \cdot  \_H,
    \quad 
    \_B = \overline{\overline{\mu}} \cdot  \_H +\overline{\overline{\zeta}} \cdot   \_E,
    \l{constitutive}
\end{equation}
the physical origin of polarisation-insensitive NDD is attributed to the synthetic moving medium effect expressed by the magnetoelectric tensors $\overline{\overline{\xi}}=\overline{\overline{\zeta}}^T$, both being antisymmetric. Here, $\_D$ and $\_B$ represent electric and magnetic displacement fields, while $\_E$ and $\_H$ are the electric and magnetic fields. 
As seen from \r{constitutive}, the synthetic motion effect manifests in the induction of high-frequency magnetisation by the electric field component of incident light, with the resulting magnetisation $\_B$ oriented orthogonally to both the electric field $\_E$ and the direction of light propagation. A similar magnetisation arises in a genuinely moving medium due to the Lorentz transformation, which motivates the nomenclature of the effect~\cite{463d18444148452ea4bc91b24b1d7c65}.

\begin{figure}[tb]
    \centering
    \includegraphics[width=0.9\linewidth]{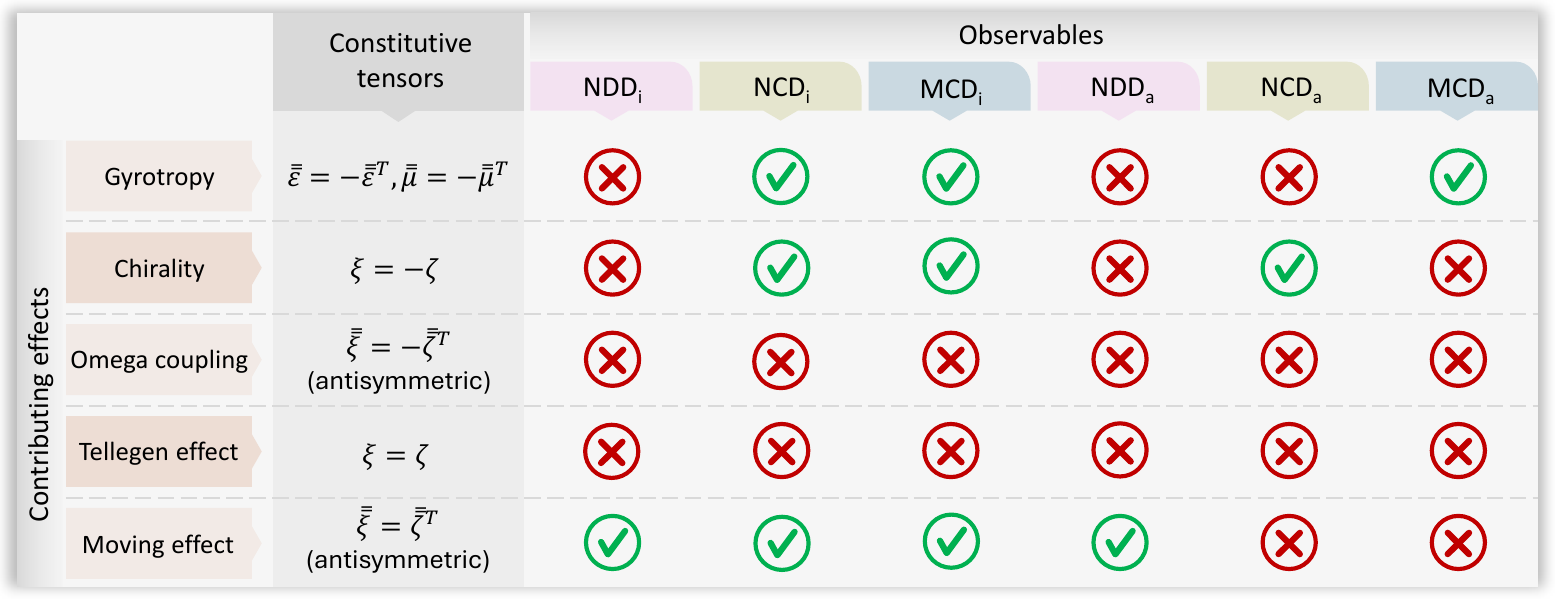}
    \caption{\textbf{Comparison of dichroism observables and the different contributing electromagnetic effects.} Green ticks and red crosses indicate whether a specific electromagnetic effect contributes to a particular observable. Here, 'NDD' stands for nonreciprocal directional dichroism, 'MCD' for magnetic circular dichroism, and 'NCD' for natural circular dichroism.
    }
    \label{fig:table}
\end{figure}

Since NDD requires the breaking of both $\mathcal{P}$ and $\mathcal{T}$ symmetries, it is instructive to compare it with two well-known circular dichroism effects that can also arise in uniaxial materials: natural circular dichroism (NCD) and magnetic circular dichroism (MCD). The former emerges from broken $\mathcal{P}$ symmetry, while the latter originates from broken $\mathcal{T}$ symmetry (see Figs.~\ref{fig:Panel1}b and c). 
Originally, NCD was defined as the contrast in light absorptivities for opposite circular polarisations. Regardless of the illumination direction, chiral materials predominantly absorb one specific circular polarisation (Fig.~\ref{fig:Panel1}b). In the case of colloids, such a definition attributed NCD solely to the chirality of the material~\cite{Barron_2004}. However, for anisotropic materials, the conventional definition of NCD becomes ambiguous, as it can yield a non-zero value even when the material is achiral~\cite{albooyeh2023classification}. 
Consequently, modern literature has introduced various parameters to quantify the level of chirality in a material~\cite{zhang2019measuring,decker2007circular,hentschel2012three,frank2013large,esposito2015nanoscale}. As a generalisation of these, we select the following intensity-based definition for our analysis $\mathrm{NCD_i}=(T_{\rm LL}^+ -T_{\rm RR}^+ +T_{\rm LL}^- -T_{\rm RR}^-)/2$,
where the indices ‘R’ and ‘L’ denote right- and left-hand circular polarisations, respectively.
Nonetheless, even this definition does not exclusively capture the chirality of a given uniaxial material system. As shown in Fig.~\ref{fig:table}, a non-zero ${\rm NCD_i}$ can result from synthetic motion effects and gyrotropic phenomena, even in the absence of intrinsic material chirality. The results summarised in the table are derived in Supplementary Section~1. We observe that only in the absence of nonreciprocal effects (i.e., no time-odd bias fields in the material) can a non-zero ${\rm NCD_i}$ be uniquely attributed to the material chirality.

Likewise, MCD—manifested as direction-dependent circular dichroism (see Fig.~\ref{fig:Panel1}c)—can be quantified by the intensity-based expression
$\mathrm{MCD_i}=(T_{\rm LL}^+ -T_{\rm RR}^+ -T_{\rm LL}^- +T_{\rm RR}^-)/2$. 
 As shown in Fig.~\ref{fig:table}, according to this definition, a non-zero $\mathrm{MCD_i}$ can arise not only from gyrotropic (Faraday) effects but also due to synthetic motion accompanied by material chirality.

To isolate the different effects illustrated in Fig.~\ref{fig:table}, we introduce dichroism observables based on the complex field \textit{amplitudes} denoted by the subscript `{\rm a}' (see derivations in Supplementary Section~1):
\begin{equation}\begin{split}
\mathrm{NDD_a}&=\left| t_{xx}^{+} - t_{xx}^{-} + t_{yy}^{+} - t_{yy}^{-}\right|^2/4,\\ 
\mathrm{NCD_a}&=\left| t_{\rm LL}^{+} - t_{\rm RR}^{+} + t_{\rm LL}^{-} - t_{\rm RR}^{-}\right|^2/4,\\
\mathrm{MCD_a}&=\left| t_{\rm LL}^{+} - t_{\rm RR}^{+} - t_{\rm LL}^{-} + t_{\rm RR}^{-}\right|^2/4.
\end{split}
\end{equation}
According to these definitions, $\mathrm{NDD_a}$, $\mathrm{NCD_a}$, and $\mathrm{MCD_a}$ uniquely quantify the synthetic moving effect, chirality, and gyrotropic effects, respectively. Although measuring them optically requires capturing both the complex amplitude and phase of the transmission coefficient (via coherent detection), these quantities enable efficient extraction and independent quantification of each effect.

The aim of this work is to design a metasurface with  
$\mathrm{NDD_a} \approx 1$ and $\mathrm{NCD_a} \approx 0$ and $\mathrm{MCD_a} \approx 0$. Such a metasurface would exhibit perfect polarisation-insensitive one-way transparency and the optical response of a purely synthetic moving medium.

\subsection*{Enhancing NDD using quasi-BICs}

To design a metasurface that exhibits maximal NDD, we first identify a unit cell with the appropriate symmetry and then optimise the structure to maximise the effect. 
According to the magnetic point group classification (Shubnikov's classification), it is known that, for bulk materials, the spatiotemporal symmetry that supports solely the artificial moving medium effect—while suppressing all other gyrotropic or bianisotropic effects—is described by the third-kind magnetic point group $D_{4 \rm h}(C_{4 \rm v})$~\cite{dmitriev_group_1999,Mackay_Lakhtakia_anisotropy_2005} (see the full list of appropriate magnetic point groups in Supplementary Section~2). This specific symmetry ensures that the magnetoelectric tensors in~\r{constitutive} are antisymmetric and satisfy the relation $\overline{\overline{\xi}} = \overline{\overline{\zeta}}^T$.
For metasurfaces, instead of employing the bulk description provided by~\r{constitutive}, we adopt a microscopic framework based on collective polarizabilities~\cite{Radi2014}.
\begin{equation}
\mathbf{p}= \overline{\overline{\alpha}}_{\rm ee} \cdot \mathbf{E}_{\rm inc} + \overline{\overline{\alpha}}_{\rm em} \cdot \mathbf{H}_{\rm  inc},    \quad 
\mathbf{m}= \overline{\overline{\alpha}}_{\rm mm} \cdot \mathbf{H}_{\rm inc} + \overline{\overline{\alpha}}_{\rm me} \cdot \mathbf{E}_{\rm inc},
    \label{pandm}
\end{equation}
where $\_p$ and $\_m$ represent the dipole moments of the unit cells, and $\_E_{\rm inc}$ and $\_H_{\rm inc}$ denote the incident electric and magnetic fields. 
When applying the specified magnetic point group to the metasurface polarizabilities (see derivations in Supplementary Section~2), we find that 
the electric and magnetic tensors must be diagonal, $\alpha_{\rm ee}^{xx}= \alpha_{\rm ee}^{yy}$ and  $\alpha_{\rm mm}^{xx}= \alpha_{\rm mm}^{yy}$, while the magnetoelectric tensors antisymmetric, $\alpha_{\rm em}^{xy}=-\alpha_{\rm em}^{yx}=-\alpha_{\rm me}^{xy}=\alpha_{\rm me}^{yx}$.

\begin{figure}[H]
    \centering
    \includegraphics[width=0.8\linewidth]{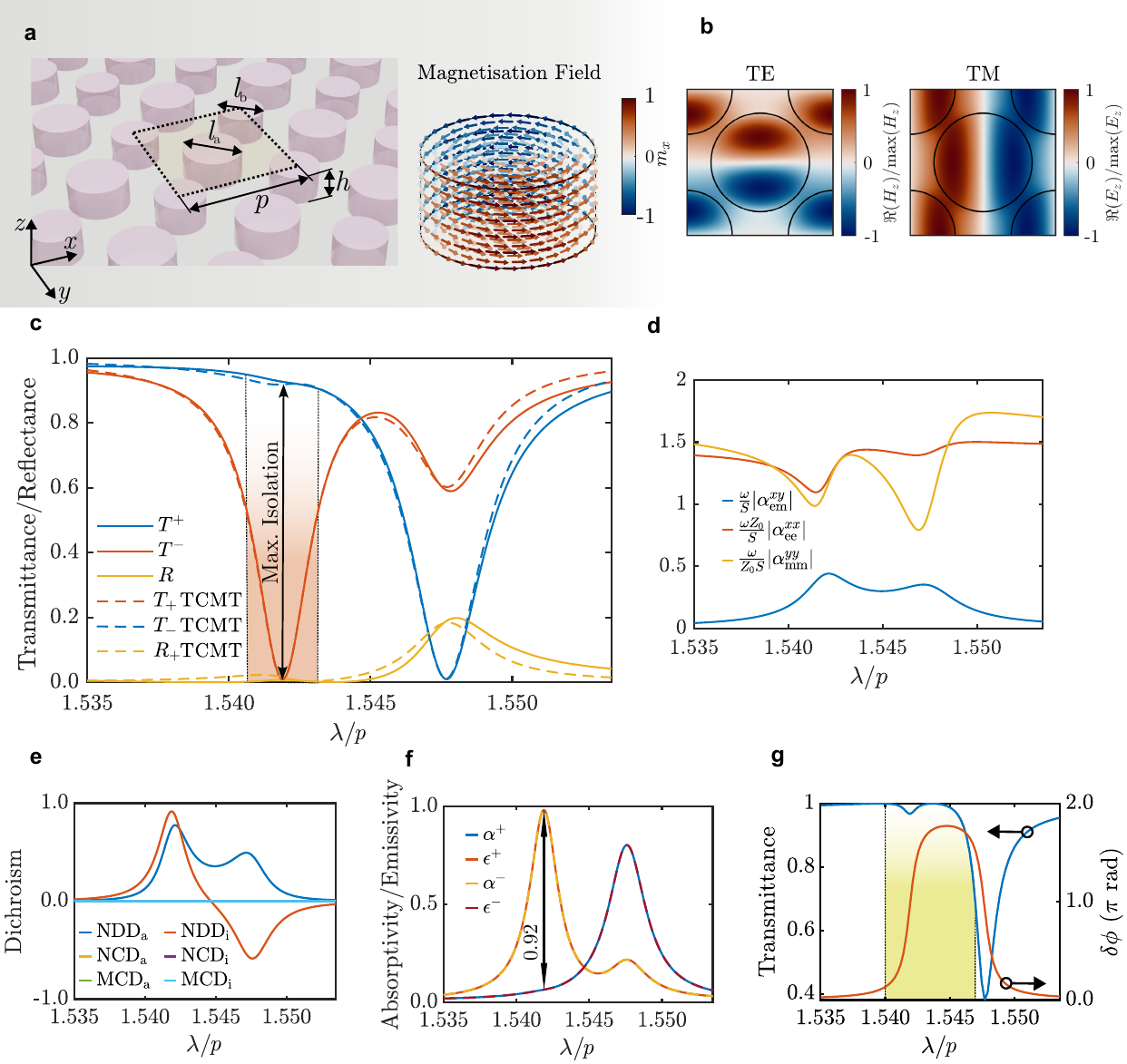}
    \caption{ \textbf{Electromagnetic response of a bias-free metasurface supporting vortex-type static magnetisation and quasi-BICs}. \textbf{a,} Metasurface geometry (left) and static magnetisation within each nanodisk (right). The dashed contour depicts the square unit cell. The neighboring nanodisks have slightly different diameters $l_{\rm a}$ and $l_{\rm b}$ and are arranged in the checkerboard order, enabling the excitation of quasi-BICs in the metasurface. The colorbar depicts the amplitude of the normalised magnetisation $\_M/M_{\rm s}$ inside the nanodisk, while the arrows show the orientation of the local magnetisation. 
    \textbf{b,} Field profiles of the quasi-BICs orthogonal eigenmodes excited in the metasurface unit cell in the absence of magnetisation. The left and right plots depict an electric (TE) and magnetic (TM) dipole modes, respectively. \textbf{c,} Scattering parameters of the metasurface consisting of idealistic ferrite material when illuminated normally with unpolarised light. Both simulated data and the theoretical data from the TCMT model are shown. The chosen ferrite parameters are dispersionless with $\varepsilon_r=8$, $\varepsilon_i=0.01$, and $\varepsilon_a=0.1$. Geometric parameters of the metasurface are, in terms of the period, $D=0.6446p$, $h=0.3261p$, and $\Delta=l_a-l_b=0.0557p$, with diameters defined as $l_a=D+\Delta/2$ and $l_b=D-\Delta/2$. The wavelength in the plot is normalised by the metasurface period $p$.
    \textbf{d,} Absolute values of the normalised collective polarizabilities of the metasurface unit cell extracted from the simulated scattering coefficients. 
    \textbf{e,} Simulated directional and circular dichroisms of the metasurface. Indices ``i'' and ``a'' correspond to dichroism quantities defined in the text in terms of intensities and field amplitudes, respectively. 
    \textbf{f,} Emissivities and absorptivities of the metasurface for two illumination directions.  \textbf{g,} Transmittance and transmission phase difference between opposite directions of propagation for the same metasurface as in \textbf{c} but in the absence of material losses ($\varepsilon_{\rm i}=0$).}
    \label{fig:Panel2}
\end{figure}

We choose as the meta-atom a ferrite nanodisk with in-plane magnetisation forming a closed circular loop around its axis of rotational symmetry. In the next section, we discuss how such a vortex magnetisation can be realised in ferrite nanodisks even in the absence of external magnetizing fields. Here, we first design a metasurface without specifying a particular ferrite material.
The nanodisks have the continuous form of the desired spatiotemporal symmetry, that is, $D_{\infty \rm h}(C_{\infty \rm v})$, as shown in Fig.~\ref{fig:Panel2}a. In the figure inset, the arrows show the direction of the normalised magnetisation $\textbf{m}=\mathbf{M}/M_{\rm s}$, where $\mathbf{M}$ is the static magnetisation and $M_{\rm s}$ is the saturation magnetisation of the nanodisk material. We distribute the nanodisks in a square lattice to form a two-dimensional periodical metasurface with $D_{4 \rm h}(C_{4 \rm v})$ symmetry. Therefore, with the aforementioned spatiotemporal symmetry, our metasurface provides a pure artificial moving medium effect. Furthermore, the aforementioned symmetry ensures polarisation-insensitive response and simplifies the analysis since it can be described as a two-port system (see Supplementary Section~3).

It is important to recognise that, although the nanodisk material exhibits only a local gyroelectric response (characterised by an anisotropic, inhomogeneous permittivity tensor $\overline{\overline{\varepsilon}} = -\overline{\overline{\varepsilon}}^T$) and lacks any bianisotropic effects, the metasurface composed of these nanodisks demonstrates an effective bianisotropic response. Specifically, it behaves as a pure artificial moving medium, exhibiting no gyrotropic responses (compensated due to vortex magnetisation), with diagonal electric and magnetic tensors and anti-symmetric magnetoelectric tensors satisfying $\overline{\overline{\alpha}}_{\rm em} = \overline{\overline{\alpha}}_{\rm me}^T$.

In order to obtain a sufficiently strong artificial moving medium effect, we introduce symmetry-protected quasi-BICs to the metasurface. Such resonances can have unlimited quality factors, which allows to strongly enhance the interaction time of light with the metasurface and therefore drastically enhance the otherwise weak gyrotropic effects in ferrites at optical frequencies. 
 For this reason,  the metasurface shown in Fig.~\ref{fig:Panel2}a features a chessboard-type meta-atom arrangement where the neighbouring nanodisks have a slight difference in diameter $\Delta$. The parameters $l_a=D+\Delta/2$ and $l_b=D-\Delta/2$ are the diameters of the larger and smaller nanodisks, being perturbed from the average diameter value $D$.  
Perturbing in this way the diameter of the nanodisks folds the first Brillouin zone, bringing a previously inaccessible state at the $X$ point to the $\Gamma$ point (normal incidence)~\cite{Overvig2020,Geometry-1,Geometry-2,Mez-Espina2024}. The spatiotemporal point symmetry of the structure remains invariant after this perturbation. Moreover, it is known that quality factor of symmetry-protected quasi-BICs follows an inverse-squared law with respect to the perturbation amplitude, that is, $Q\simeq C/\Delta^2$, where $C$ is a constant~\cite{Koshelev2018}. This adds a degree of freedom for tuning the quality factor of the metasurface.

To achieve ${\rm NDD_a} \sim 1$, it is essential to ensure complete light absorption for illumination from one direction and full transmission from the opposite direction. Both conditions require the elimination of reflection, which can only be realised in a metasurface that supports simultaneously induced effective electric and magnetic currents~\cite{Radi2014,decker2015high}. Accordingly, we select the geometric parameters of the nanodisks (listed in the caption of Figure~\ref{fig:Panel2}) such that they support overlapping electric (transverse-electric, TE-type) and magnetic (transverse-magnetic, TM-type) quasi-BIC dipolar resonant modes at the same frequency.
Figure~\ref{fig:Panel2}b illustrates the normalised $H_z$ and $E_z$ components of these two degenerate eigenmodes in the absence of static magnetisation in the nanodisks. These quasi-BIC eigenmodes correspond to the two-dimensional $E$ representation of the spatial group $C_{4 \rm v}$. By aligning the two resonances, the so-called Huygens' condition is fulfilled~\cite{decker2015high}, which—when combined with critical coupling between scattering and intrinsic losses—can suppress light reflection and enhance absorption~\cite{Radi2014,asadchy2015broadband}. 
Introducing the static magnetisation in the nanodisks, nonreciprocal bianisotropic coupling (i.e., the artificial moving medium effect) emerges, breaking the orthogonality of the electric and magnetic dipole modes (see more details in Supplementary Section~4). As a result, the spatiotemporal symmetry of the structure enforces direction-dependent scattering of light. Upon introducing magnetisation in the nanodisk, the dipolar modes hybridise; hence, we refer to them hereafter as hybrid electric and hybrid magnetic modes.

We model the scattering properties of the nonreciprocal metasurface using an original TCMT~\cite{CMT-OG,CMT-eqs,CMT-nonreciprocal}, incorporating appropriate nonreciprocal constraints (see detailed derivation in Supplementary Section~5). In our coupled-mode equations, the electric and magnetic resonant modes are coupled due to mode hybridisation induced by the static magnetisation within the nanodisks. Specifically, the vortex magnetisation breaks the out-of-plane mirror symmetry, rendering the two resonances no longer orthogonal.
Next, we derive from the TCMT model the conditions required to maximise ${\rm NDD_i}$ in the metasurface. By analyzing the transmission contrast at one of the resonance frequencies, we conclude that ${\rm NDD_i}$ reaches its maximum when the following three conditions are satisfied (see details in Supplementary Section~5): (i) the scattering and absorption decay rates must be equal for each resonant mode (the so-called critical coupling condition~\cite{CMT-Absorption-condition-2}); (ii) the magnetic and electric resonances must spectrally overlap (the so-called Huygens' condition); and (iii) the coupling between resonant states must be maximised and/or the quality factors of the resonances must be high.
If these conditions are met, and the possible substrate on which the metasurface resides is matched to free space (such that the metasurface is fully transparent at wavelengths far from the resonances), it can be shown that at one of the resonant wavelengths, the maximum value of directional dichroism is given by
\begin{equation}\label{eq:Delta_T_CMT}
    \mathrm{NDD_i}=T^+-T^-=\frac{g^2}{\gamma^2+g^2},
\end{equation}
where $g$ is the coefficient responsible for the coupling between the two resonant modes, and $\gamma$ is the decay rate of both resonances (here, the scattering and absorption decay rates are equal due to the critical coupling condition). In practice, the coupling coefficient $g$ is determined by the permittivity tensor of the ferrite material and is proportional to the magnitude of its off-diagonal component (see Supplementary Section~5). In the optical regime, this component is typically very small, implying that $g$ is also a small quantity. Nevertheless, as evident from Eq.~(\ref{eq:Delta_T_CMT}), it is still possible to achieve high $\mathrm{NDD_i}$ values if the decay rate $\gamma$ is engineered to be sufficiently small, specifically when $\gamma \ll g$. Here, we employ quasi-BICs that allow us to achieve arbitrarily small values of $\gamma$ by tuning the perturbation factor $\Delta$, as $\gamma \sim \Delta^2$.

We carry out full-wave simulations of the nonreciprocal metasurface with the geometric parameters optimised via TCMT (mentioned in the caption of Fig.~\ref{fig:Panel2}). Note that in the plots in Fig.~\ref{fig:Panel2}, we normalise all the geometric metasurface parameters and the wavelength to the metasurface period $p$. This conventional normalisation allows straightforward scalability of all the design parameters for any operating wavelength (assuming that the permittivity tensor remains the same). 
For the simulation, we first adopt generic material parameters for the ferrite nanodisks, chosen to approximate typical values for conventional ferrite materials such as YIG or BIG (Bi3YIG)~\cite{Meta-BiYIG-material}. Although now we focus on a simplified proof-of-principle design, below we also examine a practical metasurface incorporating specific ferrite material capable of supporting vortex-type magnetisation and accounting for various material imperfections. 

The space-nonuniform permittivity tensor of generic ferrite is modeled to first order as follows~\cite{zvezdin1997modern}:
\begin{equation}
\overline{\overline{\varepsilon}} (\_r)= \frac{\varepsilon_0}{M_{\rm s}}
\begin{pmatrix}
\varepsilon_{\rm r}-j \varepsilon_{\rm i} &    j \varepsilon_{\rm a} M_z(\_r) &  - j\varepsilon_{\rm a} M_y(\_r)  \\
 - j\varepsilon_{\rm a} M_z(\_r)  & \varepsilon_{\rm r}-j \varepsilon_{\rm i} &    j\varepsilon_{\rm a} M_x(\_r)  \\
   j\varepsilon_{\rm a} M_y(\_r)  &  - j\varepsilon_{\rm a} M_x(\_r)  & \varepsilon_{\rm r}-j \varepsilon_{\rm i}
\end{pmatrix},
\end{equation}
where $\varepsilon_{\rm r}$ and $\varepsilon_{\rm i}$ are the real and imaginary parts of the diagonal component of the relative permittivity, respectively, $\varepsilon_{\rm a}$ is the off-diagonal permittivity component due to the static magnetisation, and $M_i$ are the components of the local vortex magnetisation that are functions of the spatial coordinate $\_r$. The magnetisation vector field has an in-plane distribution as shown in Fig.~\ref{fig:Panel2}a. More details about the simulations can be found in Methods. 
Due to the symmetry of the structure, the only non-zero components of reflectances and transmittance are $T_{xx}^+=T_{yy}^+=T^+$, $T_{xx}^-=T_{yy}^-=T^-$, and $R_{xx}^+=R_{yy}^+=R_{xx}^-=R_{yy}^-=R$ (see Supplementary Section~3). The simulated scattering parameters are plotted in Fig.~\ref{fig:Panel2}c together with the parameters predicted by our TCMT model.  The fitted TCMT parameters for the metasurface are provided in Supplementary Section~5. The figure reveals two spectrally separated dips in transmittance for oppositely directed incident light, indicating a giant directional dichroism exhibited by the optical metasurface. At the normalised wavelength $\lambda/p = 1.542$, the polarisation-insensitive isolation reaches $10 \log_{10} (T^+ / T^-) = 19.71$ dB. At this wavelength, $92.68\%$ of the incident energy is transmitted in the forward direction, whereas only $0.99\%$ is transmitted in the opposite direction. Additionally, the reflectance for both illumination directions remains below $0.8\%$. Notably, this substantial directional dichroism is achieved despite the sub-wavelength thickness of the metasurface.
The TCMT model demonstrates excellent agreement with the simulated results. Minor discrepancies between the two are attributed to the assumption of a frequency-independent scattering matrix for the background channel and the finite decay rates. Figure~\ref{fig:Panel2}e depicts ${\rm NDD_i}$ and ${\rm NDD_a}$  calculated from the simulated scattering parameters of the metasurface. The ${\rm NDD_i}$ reaches $0.919$, while ${\rm NDD_a}$ is as high as $0.777$, while NCD and MCD are negligibly small in the considered wavelength range. 

To show that the metasurface exhibits the effect of an artificial moving medium, we plot in Fig.~\ref{fig:Panel2}d   its main in-plane collective polarizabilities extracted from the scattering parameters of the metasurface (see Supplementary Section~1). In the plot, the absolute values of the normalised polarizabilities are shown. The collective polarizabilities indeed have a pure moving signature expressed as $\alpha_{\rm em}^{xy}=-\alpha_{\rm em}^{yx}=-\alpha_{\rm me}^{xy}=\alpha_{\rm me}^{yx}$, while $\alpha_{\rm ee}^{xx}=\alpha_{\rm ee}^{yy}$ and $\alpha_{\rm mm}^{xx}=\alpha_{\rm mm}^{yy}$. The magnetoelectric polarizability exhibits resonant peaks that correspond to the resonant wavelengths of the hybrid electric and magnetic modes within the nanodisks. As is seen in Fig.~\ref{fig:Panel2}d, the normalised magnetoelectric (moving effect) polarizability is of the same order of magnitude as the electric and magnetic polarizability components. This is a noteworthy result at optical frequencies, where magnetoelectric and gyrotropic effects in the vast majority of materials are many orders of magnitude smaller than the corresponding permittivity or permeability values.

Figure~\ref{fig:Panel2}f presents the emissivities $\epsilon^\pm$ and absorptivities $\alpha^\pm$ of the metasurface under illumination from both directions, calculated from the scattering parameter spectra (see Supplementary Section~6 for details). Owing to the absence of Lorentz reciprocity, the emissivity and absorptivity are unequal for a given illumination direction, i.e., $\epsilon^+ \neq \alpha^+$ and $\epsilon^- \neq \alpha^-$, in contrast to reciprocal metasurfaces where Kirchhoff’s law of radiation ensures their equality. At the normalised wavelength $\lambda/a = 1.542$, the contrast between emissivity and absorptivity reaches a maximum of $\epsilon^+ - \alpha^+ = \alpha^- - \epsilon^- \simeq 0.92$. 
This pronounced contrast is particularly crucial for the design of nonreciprocal thermal systems aimed at enhancing energy harvesting in solar cells. Although recent studies have also reported significant emissivity-absorptivity contrast~\cite{Park2021,CMT-Absorption-condition-2,park2024kirchhoff}, they were restricted to a single light polarisation, whereas the presented metasurface functions effectively under unpolarised light.

Another important application of the proposed metasurface arises when it operates in a nearly lossless regime. In this case, the transmittance for light incident from opposite directions must be equal (see Supplementary Section~3); however, the complex phase of the transmission coefficient may differ. 
Such an electromagnetic response enables the realisation of a photonic gyrator—the so-called fifth linear element, in addition to the resistor, capacitor, inductor, and transformer—offering a new degree of freedom for synthesizing arbitrarily complex optical circuits~\cite{Yang2023}. 
Figure~\ref{fig:Panel2}g shows the transmittance $|T^+| = |T^-|$ and the phase difference $\delta\phi = \angle t_+ - \angle t_-$ for the same metasurface geometry as in Fig.~\ref{fig:Panel2}a, with the only modification being the assumption of a lossless response ($\varepsilon_{\rm i} = 0$). 
Near the wavelength $\lambda/a = 1.542$, the transmittance exceeds 95\% while the phase difference spans nearly the full $2\pi$ range. This wide phase variation opens up exciting possibilities for nonreciprocal (direction-dependent) beam steering~\cite{Yang2023} at optical frequencies. Notably, this remarkable behavior is achieved without any optimisation and thus offers potential for further enhancement.

\subsection*{Uniform arrays with magnetic vortices}

Here, we discuss the practical realisation of the metasurface with magnetic vortices depicted in Fig.~\ref{fig:Panel2}a using realistic ferrite materials. Specifically, we consider bismuth iron garnet  ${\rm Bi_3Fe_5O_{12}}$ (abbreviated as BIG), due to its strong magneto-optical response~\cite{Meta-BiYIG-material}. A similar analysis for the widely used yttrium iron garnet (YIG) is provided in Supplementary Section~7.

\begin{figure}[h]
    \centering
    \includegraphics[width=0.85\linewidth]{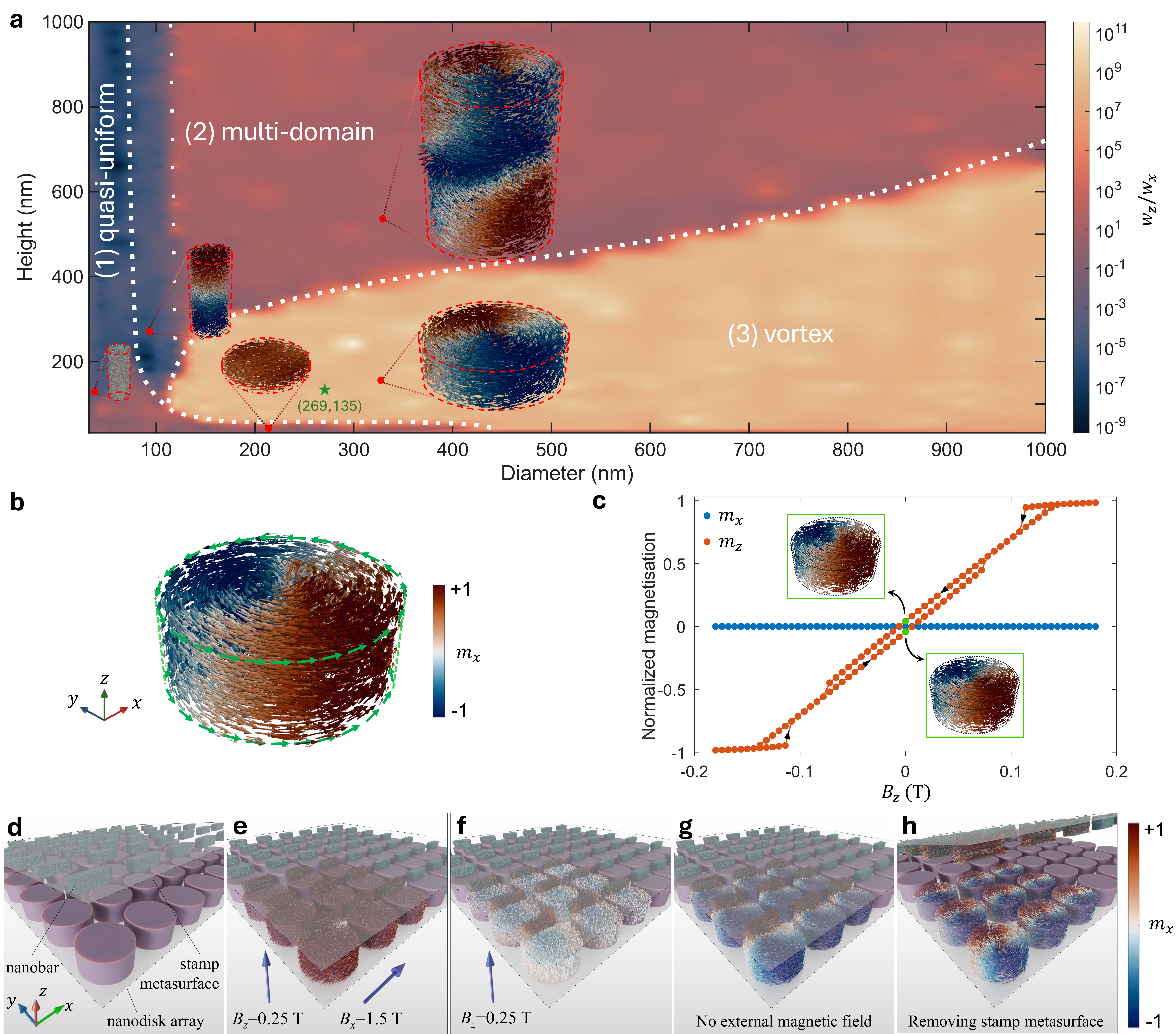}
        \caption{ \textbf{Micromagnetic simulations of BIG nanodisks}. \textbf{a,} Phase diagram of magnetic states in a single BIG nanodisk for different values of its diameter $D$ and height $h$. The color denotes the ratio between the $z$- and $x$-components of the vorticity on a logarithmic scale.  A larger ratio corresponds to a greater in-plane curl and therefore a more vortex-like state. The white dashed lines depict the approximate boundaries between different magnetic states in the diagram. The insets depict principal states, where the arrows depict the directions of normalised local magnetisation $\_m=\_M/M_{\rm s}$ and the arrow color denotes the $m_x$ component. The star refers to the point with optimised dimensions of the nanodisks for the metasurface design. 
        \textbf{b,}  Spatial distribution of the normalised magnetisation component $m_x$ in the single nanodisk with optimised parameters, $D=269$~nm and $h=135$~nm, corresponding to the point marked by the star in \textbf{a}. The nanodisk requires no external magnetisation after the vortex state is established. 
        \textbf{c,} Hysteresis plot for the nanodisk in \textbf{b} for varying applied external magnetic field oriented along the $z$-direction. \textbf{d-h,} An original methodology for generating identical vortex states in nanodisk arrays. Here, the nanodisks have the same dimensions as the nanodisk in \textbf{b}. The method is applied to an initial $3 \times 3$ array using a micromagnetic simulator, with magnetisation configurations extracted at each step. The colormap shows the normalised magnetisation component $m_x$ in the nanodisks and nanobars at different stages of the proposed magnetisation procedure. The array of rectangular bars above the nanodisk array represents an auxiliary stamp metasurface.}
    
    \label{BIG_mic}
\end{figure}

First, we perform rigorous micromagnetic simulations on a single BIG nanodisk to determine the conditions under which it can sustain a stable magnetic vortex state. Subsequently, we examine periodic arrays of such nanodisks and demonstrate how to ensure that they exhibit identical vortex states despite their mutual magnetic interactions. The details of the micromagnetic simulations and the magnetic parameters of BIG can be found in Methods. 
In Fig.~\ref{BIG_mic}a, we plot a phase diagram of stable types of magnetic states in nanodisks depending on their diameter $D$ and height $h$. 
To construct this diagram, we introduced the magnetic vorticity defined as
\begin{equation}
     \boldsymbol{w} = \frac{1}{M_{\rm s} V} \int_V \nabla \times \mathbf{M} \, dV,
\end{equation}
where \( V \) denotes the nanodisk volume, and the integration is performed over this volume.  The color in the phase diagram depicts the ratio \( w_z / w_x \), which quantitatively characterises the vortex nature of the magnetisation in the desired $z$-direction (see the right panel in Fig.~\ref{fig:Panel2}a). The normalisation of $w_z$ by $w_x$ enables a clearer comparison between helical and vortex magnetisation configurations, as in the case of YIG nanodisks (see Supplementary Section~7). In both configurations, $w_z$ is large, while $w_x$ is comparatively smaller for the vortex state. However, this normalisation is not essential here, as plotting $w_z$ alone yields a similar phase map owing to the absence of a helical magnetisation state.

To generate the diagram, we applied a strong static external field $B_z$ for saturating the nanodisk and then reduced it to zero for observing the final magnetisation state. 
Within the explored parameter space of \( D \) and \( h \), magnetisation configurations can be broadly classified into three principal states labeled and separated by white dashed lines in Fig.~\ref{BIG_mic}a: (1) quasi-uniform, (2) multi-domain, and (3) vortex. 
For very narrow ($D < 50$~nm) or very thin ($h < 50$~nm) cylinders, the magnetisation is predominantly quasi-uniform. In rod-like geometries, the magnetisation aligns along the $z$-axis, while in disk-like structures, it lies in-plane (see the corresponding insets in Fig.~\ref{BIG_mic}a). 
Two distinct types of multi-domain states are identified, corresponding to the different color regions in the phase diagram (see the insets in the figure). For cylinders with smaller diameters ($60$~nm$ < D < 120$~nm), the ground state comprises multiple quasi-uniform domains oriented in various directions. Cylinders with larger diameters ($130$~nm$< D$) exhibit a two-domain state characterised by oppositely curling magnetisations.

Vortex states, as illustrated in the inset of Fig.~\ref{BIG_mic}a, constitute the ground state over a wide range of $D$ and $h$. A vortex state is uniquely characterised by two binary parameters: circularity, which denotes the direction of in-plane magnetisation circulation—counterclockwise ($c = +1$) or clockwise ($c = -1$)—and polarity, which indicates the orientation of the vortex core magnetisation perpendicular to the nanodisk plane—upward ($p = +1$) or downward ($p = -1$). 
Thus, the magnitude of the ratio \( w_z / w_x \) serves as an effective metric to differentiate among vortex and other types of magnetic states.

We analyse in more detail the vortex state in a single nanodisk.  Figure~\ref{BIG_mic}b shows the normalised magnetisation direction $\_m=\_M/M_{\rm s}$ (the color denotes the $m_x$ component) within a nanodisk with dimensions optimised in the next section for supporting Huygens'-type quasi-BIC resonances ($D = 269$~nm and $h = 135$~nm). The magnetisation configuration exhibits a nearly-perfect magnetic vortex state with $c = +1$ and $p = +1$. 
Experimental generation of such a vortex state in a single nanodisk is straightforward: one needs to apply a strong magnetic field along the $z$-axis to saturate the magnetisation in that direction, followed by a gradual reduction of the external field to zero. The hysteresis loop for such magnetisation of the nanodisk is shown in Fig.~\ref{BIG_mic}c. 
The $m_x$-component remains zero while $m_z$ varies with $B_z$. The insets in the figure correspond to the two green points in the hysteresis at $B_z=0$ where vortex states with opposite polarities and identical circularities are established. 
At $B_z = 0$, $m_z$ remains non-zero due to the vortex polarity, although its magnitude is very small.  The presence of a finite $m_z< 0.05$ introduces additional, although weak, optical effects to our metasurface, such as Faraday rotation.  A detailed analysis of these effects is presented in Supplementary Section~8. Note that vortex states in single ferrite nanodisks were experimentally achieved in previous works~\cite{losby2015torque,barron2021magnetization,collet2016generation,zhu2017patterned}.

To achieve strong NDD in our metasurface, it is essential that all nanodisks support magnetic vortices with identical circularity; in other words, the metasurface must comprise a uniform array of magnetic vortices. Although the challenge of generating uniform vortex arrays has been explored in several experimental studies~\cite{im2017simultaneous,jaafar2010control,im2012symmetry,pradhan2023control,giesen2007vortex}, none of these approaches are compatible with our design. These methods rely on breaking $C_4$ symmetry, which is not suitable here.
A configuration consisting of identical vortices in an array of magnetic nanodisks (or other nanoparticle geometries) possesses higher potential energy than one with approximately equal populations of clockwise and counterclockwise vortices, resulting in nearly zero net circularity (i.e., antiferromagnetic order). Consequently, simple magnetisation protocols would fail to realise uniform-vortex states in a proposed metasurface.
Nevertheless, our micromagnetic simulations confirm that this uniform-vortex state is stable—once established, it persists indefinitely in the absence of significant external perturbations.

We thus introduce a deterministic method for generating vortices with uniform circularity in an array of nanodisks. This approach can be readily extended to magnetic arrays exhibiting non-uniform but arbitrarily prescribed vortex distributions, as well as to metasurfaces composed of elements with alternative geometries and materials. 
The method is based on the use of a removable auxiliary magnetic metasurface—referred to as the stamp metasurface. This auxiliary metasurface comprises an array of ferromagnetic rectangular nanobars with the same periodicity as the target array of the nanodisks. The nanobars are oriented parallel to the \( x \)-axis. Fabrication of such metasurfaces with ferromagnetic patterns has been an established process~\cite{maccaferri2016anisotropic, wang2020probing}.
In the micromagnetic simulations, each nanobar is modeled with dimensions of $180\,\text{nm} \times 60\,\text{nm} \times 60\,\text{nm}$. However, these specific dimensions are not strict with slight size variations allowed. Additionally, we verified that nanowires can be employed in place of nanobars within the stamp metasurface. Cobalt is used as the ferromagnetic material in the simulations, although it is not the exclusive choice; any ferromagnetic material with strong magnetocrystalline anisotropy and saturation magnetisation may be suitable. Furthermore, this method remains effective even when the magnetocrystalline anisotropy is reduced to $40\%$ of the value for the cobalt in the hexagonal close-packed (hcp) phase, a condition that can be realised through annealing. The magnetic properties of cobalt at room temperature can be found in Methods.
Our method involves the following steps:

\begin{enumerate}
    \item \textbf{Stamp alignment:} The stamp metasurface is placed above the nanodisk array such that each nanobar is approximately located on the edge of the nanodisk, see Fig.~\ref{BIG_mic}d. Precise alignment is not critical, as simulations demonstrate that uniform vortex configurations are maintained despite minor misalignments on the order of tens of nanometers. Therefore, this step can be reliably performed using established stamp alignment techniques~\cite{yoon2017fabrication,ghahremani20243d,mcphillimy2020automated}. While direct contact is ideal, vertical separations of up to several tens of nanometers still result in vortex uniformity exceeding 90\%.

    \item \textbf{Initial saturation:}   An external magnetic field $\mathbf{B}_{\text{ext}} = (B_x, 0, B_z)$ is applied to this double-layer metasurface system. The value of $B_x$ is chosen sufficiently strong to saturate the magnetisation of both the nanodisks and the nanobars along the $x$-axis, while $B_z$ is selected to be strong enough to saturate the magnetisation of the nanodisks along the $z$-axis, but not so strong as to affect the magnetisation of the nanobars. In the micromagnetic simulations, the applied magnetic field is set to $\mathbf{B}_{\text{ext}} = (1.5, 0, 0.25)\, \mathrm{T}$, resulting in the magnetisation profile of the metasurfaces shown in Fig.~\ref{BIG_mic}e. The simultaneous application of external magnetic fields along multiple directions is experimentally viable and has been demonstrated in previous studies~\cite{harada2008direction,zhong2021simultaneous}

    \item \textbf{Field reorientation:} Setting \( B_x = 0 \) results in an external magnetic field \( \mathbf{B}_{\text{ext}} = (0, 0, B_z) \), which saturates the magnetisation of the nanodisks along the \( +z \) axis, while the nanobars retain their in-plane magnetisation along the \( +x \) axis due to the pronounced shape anisotropy of the nanobars along this axis. The corresponding magnetisation configuration is shown in Fig.~\ref{BIG_mic}f.

    \item \textbf{Gradual field reduction:} the external magnetic field is gradually reduced to zero. Meanwhile, the nanobars act as uniform magnetic dipoles aligned along the $+x$ direction due to their strong shape and magnetocrystalline anisotropies. These dipoles generate stray fields that locally modulate the vortex nucleation process in the underlying nanodisks, thereby preferentially stabilising a specific vortex circularity in each nanodisk. Consequently, a uniform circularity is established across the array of nanodisks, as shown in Fig.~\ref{BIG_mic}g. Here, the gradual reduction of the field is essential as the uniform array of vortices cannot be achieved by abruptly removing the magnetic field. If a non-uniform vortex distribution is desired, one can employ a custom-designed metasurface stamp with non-uniformly positioned $x$-oriented nanobars. These nanobars should be aligned with specific edges of the nanodisks: alignment with the positive-$y$ edge induces a clockwise vortex state, while alignment with the negative-$y$ edge results in a counterclockwise vortex state.

    \item \textbf{Stamp removal:}
     After the stamp metasurface is removed, the nanodisks retain their uniform vortex configurations as the magnetic ground state, attributed to the topological stability of the vortices. The resulting magnetisation configuration of the BIG nanodisk array is presented in Fig.~\ref{BIG_mic}h.
\end{enumerate}

We have verified through simulations that this approach consistently achieves $100\%$ uniformity across various nanodisk array sizes. For instance, see Fig.~\ref{fig:Fig_5} for the case of $5 \times 5$ BIG nanodisk arrays. This indicates its applicability to arrays of arbitrarily large dimensions, constrained only by the sizes of the fabricated target and stamp metasurfaces. The scalable nature of this method, combined with its capability to generate magnetic arrays with arbitrary, prescribed non-uniform vortex distributions, potentially makes it a highly versatile experimental tool for a broad range of applications in micromagnetics, magneto-optics, and spintronics.

\subsection*{Giant NDD in metasurface with BIG nanodisks}

\begin{figure}[t]
    \centering
    \includegraphics[width=1\linewidth]{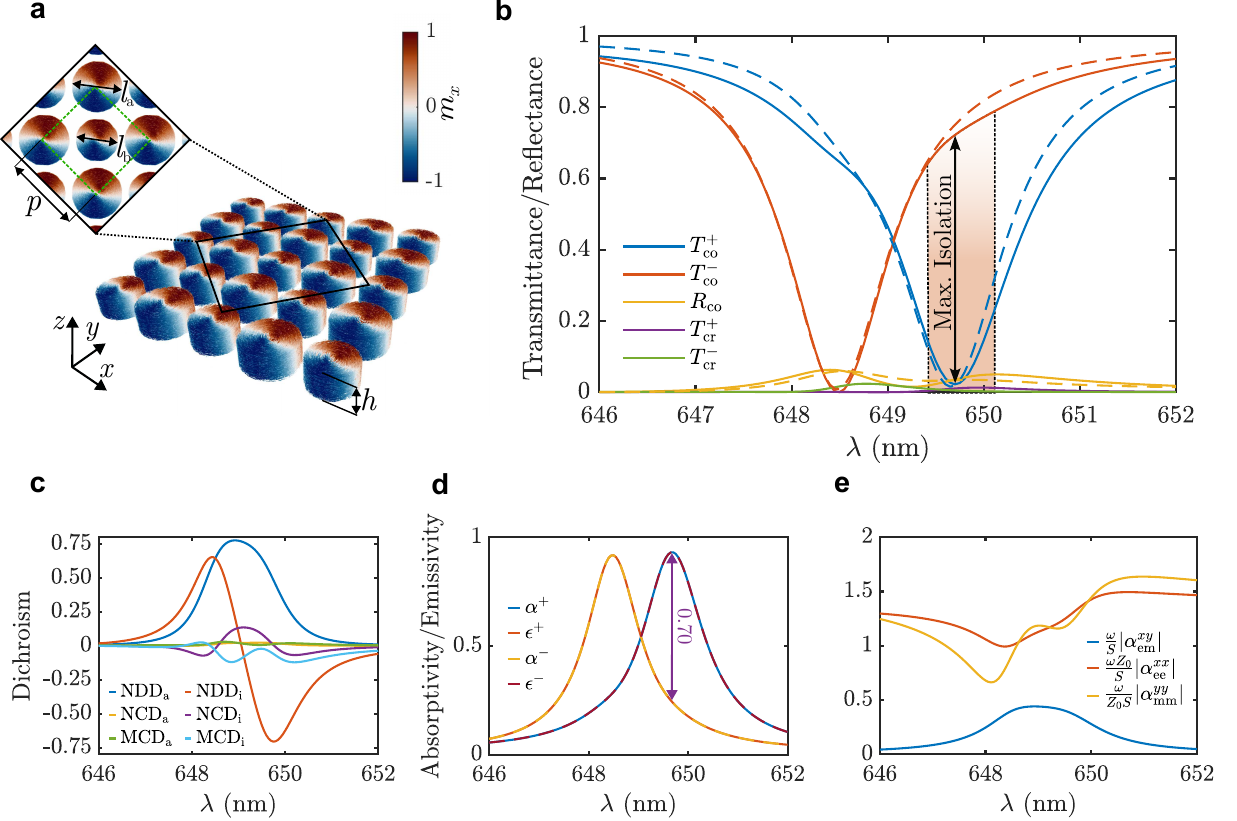}
    \caption{\textbf{Self-magnetised metasurface with BIG nanodisks.} \textbf{a,} Magnetisation field for a $5 \times 5 $ array of BIG nanodisks with the same unit cell configuration as in Fig.~\ref{fig:Panel2}a. The nanodisks require no external magnetisation. The arrows depict the directions of normalised local magnetisation $\_m=\_M/M_{\rm s}$ and the arrow color denotes the $m_x$ component. The green dashed contour depicts the unit cell. 
    Geometric parameters of the metasurface are $p=435$~nm, $D=269$~nm, $h=135$~nm, $\Delta=25$~nm. Material parameters in the considered wavelength range: $\varepsilon_{\rm r} = 8.077-0.016j$, $\varepsilon_{\rm a} = 0.0733-0.007j$~\cite{Meta-BiYIG-material}. 
    \textbf{b,} Scattering parameters of the metasurface consisting of BIG nanodisks when illuminated normally with unpolarised light. Both simulated data and the theoretical data from the TCMT model are shown. Subscripts ``co'' and ``cr'' denote co- and cross-polarised coefficients in the linear polarisation basis.
    \textbf{c,} Simulated directional and circular dichroisms of the metasurface. \textbf{d,} Emissivities and absorptivities of the metasurface for two illumination directions.
    \textbf{e,} Absolute values of the normalised collective polarizabilities of the metasurface unit cell extracted from the simulated scattering coefficients.
    }
    \label{fig:Fig_5}
\end{figure}

Since the vortex state serves as the ground state for BIG nanodisks across a wide range of geometric parameters (see Fig.~\ref{BIG_mic}a), it is straightforward to identify configurations that ensure the overlap of electric and magnetic dipolar resonant modes within the nanodisk. For our final design, we select a nanodisk with dimensions $D = 269\,\text{nm}$ and $h = 135\,\text{nm}$, whose simulated magnetic configuration is shown in Fig.~\ref{BIG_mic}b. Figure~\ref{fig:Fig_5}a presents the simulation of a $5 \times 5$ array of these nanodisks with a slight $\Delta$ perturbation to facilitate the formation of quasi-BICs, confirming uniform vorticity across the metasurface. Using the TCMT model (see Supplementary Section~5), we optimised the array's geometric parameters to meet the previously defined conditions for maximizing NDD. The material parameters and the optimised geometric parameters of the BIG nanodisks are provided in the caption of Fig.~\ref{fig:Fig_5}.

Figure~\ref{fig:Fig_5}b shows the scattering coefficients from the optimised metasurface. At the resonance frequency of $\lambda=649.75$~nm, the contrast between transmittances for opposite illuminations reaches $\Delta T=T_{\rm co}^- -T_{\rm co}^+=0.71$, shown in the plot as maximum isolation, while reflectance is kept near zero. Indeed, transmittance in the backward direction for co-polarised components is $73.18\%$, while in the backward direction, it is only $2.60\%$, and reflectance is kept under $3.88\%$. The isolation ratio reaches $14.50$~dB. The scattering parameters greatly resemble the results shown in Fig.~\ref{fig:Panel2}c. The characteristic feature of realistic vortices is that cross-polarised fields appear in transmission due to the non-zero out-of-plane magnetisation (see Supplementary Section~8). However, their amplitudes are very weak, remaining below $1.15\%$ at the resonant wavelength $\lambda_0=649.75$~nm. Conversely, at the resonant wavelength of the backwards propagating wave at $\lambda=648.5$~nm, the difference between transmittances is $\Delta T=0.65$ and reflectance of copolarised components is kept at $6.10\%$. Interestingly, the value of $T^+_{\rm co}$ decreases greatly, granting an isolation ratio of $31.5$~dB. 

From Fig.~\ref{fig:Fig_5}c, one can see that weak $\mathrm{NCD_i}$ and $\mathrm{MCD_i}$ arise from the perturbation introduced by the out-of-plane component of the magnetisation. At the resonant wavelength of $T^+_{\rm co}$, $\mathrm{NDD_i}$ reaches an impressive $-0.71$. Using the amplitude-defined dichroisms introduced earlier, it can be seen that the most prominent effect present in the metasurface is $\mathrm{NDD_a}$ (i.e. synthetic moving medium effect). While $\mathrm{NCD_a}$ and $\mathrm{MCD_a}$ are almost zero. A more detailed analysis of the dichroisms and the effects introduced by the out-of-plane component of the magnetisation can be found in Supplementary Section~8. 
In addition, Fig.~\ref{fig:Fig_5}d shows the extracted values of absorptivities and emissivities for the opposite directions of light illumination.  Their contrast reaches $\epsilon^+-\alpha^+\simeq\epsilon^--\alpha^-\simeq0.70$ at the resonant wavelength, $\lambda_0=649.75$~nm. 
Finally, we plot the calculated collective polarizabilities of the metasurface in Fig.~\ref{fig:Fig_5}e. The curves show resonant features at both resonant wavelengths. Moreover, the metasurface presents the artificial moving medium signature $\alpha_{\rm em}^{xy}=-\alpha_{\rm em}^{yx}=-\alpha_{\rm me}^{xy}=\alpha_{\rm me}^{yx}$. Nevertheless, due to the lower symmetry of the metasurface with realistic vortices, chiral and gyrotropic effects emerge; however, their polarizabilities are at least one order of magnitude weaker than the polarizabilities plotted in Fig.~\ref{fig:Fig_5}e (see Supplementary Section~8).

\section*{Discussion}

Our work merges micromagnetics, bianisotropic electrodynamics, and symmetry-protected quasi-BICs to realise sub-wavelength-thin, fully passive metasurface that approaches ideal one-way transparency for unpolarised light. By embedding self-magnetised ferrite vortices into a Huygens-type quasi-BIC platform, we obtain record-high nonreciprocal directional dichroism, yet without any external biasing field. Indeed, for our proposed metasurface design, the conventionally defined NDD parameter $\Delta \alpha_{\rm NDD}$ reaches value as high as $1.78\cdot 10^5$ ${\rm cm}^{-1}$, which exceeds those of bulk nonreciprocal media by three orders of magnitude~\cite{Park2022non}. The same design principles are transferable to other ferrites; a parallel analysis for YIG nanodisks is provided in Supplementary Section~9, underscoring the material versatility of the concept across both visible and infrared ranges.

For clarity of the physical discussion, we deliberately omitted a substrate in our analysis. Integrating one is a straightforward engineering step: the metasurface can be symmetrically embedded in a low-index host material on both sides~\cite{Yang2020}, or re-optimised on a single supporting substrate. In the latter case, the required quasi-BIC symmetry can be effectively restored by adding a thin, high-permittivity capping layer, following the strategy proposed in~\cite{bai2025recovery}. These routes preserve the key symmetry constraints that underpin both the pure synthetic moving-medium response and the high-$Q$ resonances.

An additional important outcome of this study is the deterministic, stamp-assisted protocol for writing arbitrary vortex configurations into dense nanoparticle arrays. Beyond enabling the uniform circularity needed here, this scalable and fabrication-tolerant method opens opportunities in spintronics and magneto-optics where complex, space-variant topological textures are desirable—for example, for reconfigurable magnonic crystals, programmable nonreciprocal pixels, or metasurface-based magnetic memories.

Functionally, the demonstrated platform is not limited to isolation. The analysed lossless metasurface regime with a full $2\pi$ direction-dependent phase span points to the possibility of creating bias-free photonic gyrators and nonreciprocal wavefront shapers (lenses, holograms, beam splitters, etc.). Moreover, because the structure is nearly transparent away from resonance (see Fig.~\ref{fig:Fig_5}b), multiple metasurfaces can be cascaded to broaden the operational band for thermal management, thermophotovoltaics, or radiative cooling, or even to create multi-wavelength (``RGB'') optical isolators for display and AR/VR technologies.

Taken together, these results chart a practical route toward compact, passive, and polarisation-insensitive nonreciprocal photonics. By combining high-$Q$ metasurface engineering with topologically robust magnetic textures, we provide both a device concept and a fabrication methodology that can be directly leveraged in energy harvesting, thermal photonics, and integrated nonreciprocal optics, without the burden of external magnetic bias.

\section*{Methods}
\subsection*{Full-wave optical simulations}
Electromagnetic simulations were conducted using the commercial software COMSOL Multiphysics (Electromagnetic Waves, Frequency Domain module). The software facilitates the definition of a full permittivity tensor, including position-dependent tensor components. 

In the proof of concept section, the permittivity tensor of nanodisks with ideal vortices was modelled as
\begin{equation}
\overline{\overline{\varepsilon}}(\theta) =
\varepsilon_0 \begin{pmatrix}
\varepsilon_{\rm r}-j\varepsilon_i &   0 &  - j \varepsilon_{\rm a}\cos{(\theta)} \\
 0 & \varepsilon_{\rm r}-j\varepsilon_i &  - j \varepsilon_{\rm a}\sin{(\theta)} \\
 + j \varepsilon_{\rm a}\cos{(\theta)} &  + j \varepsilon_{\rm a}\sin{(\theta)}  & \varepsilon_{\rm r}-j\varepsilon_i
\end{pmatrix}
\label{eq: Permittivity tensor comsol}
\end{equation}
This tensor form is obtained from Eq.~(5) by taking into account that an ideal vortex has zero $M_z$ magnetisation component. The magnetisation field of such ideal vortex with +1 circularity can be defined as $\mathbf{m}=\mathbf{M}/M_{\rm s}=[-\sin(\theta), \cos(\theta),0]$, where $\theta=\arctan(y/x)$ is the polar coordinate. 

However, it should be noted that in the centre of the nanodisk ($\mathbf{r} = 0$), the tensor components become ill-defined (consequence of employing cylindrical coordinates), which yields an undefined direction at the origin. To circumvent this, we used a formulation that preserves point symmetries but avoids defining $\mathbf{M}(\_r)$ at the coordinate origin. Each component of $\mathbf{M}$ was multiplied by a ramp function defined as ${\rm ramp}(2\sqrt{x^2+y^2}/l_i)$, where $l_i$, $i \in \{a, b\}$, denotes the diameter of the nanodisks (see Fig.~3a). The ramp function parameters introduced in COMSOL were: location = 0.03, slope = 9, cutoff = 1, and baseline = 0. This is translated into a one-dimensional piece-wise function that has zero value for an argument smaller than 0.03, grows linearly with a slope of 9 units for values between 0.03 and 0.14 until it reaches 1. For bigger input values, its output is always one. This adjustment ensures numerical stability by avoiding the function $\overline{\overline{\varepsilon}}$ to be ill-defined near the axis of the nanodisk.

Scattering parameter simulations were conducted using four ports, one for each polarisation on either side—as shown in Supplementary Section~3. These ports were defined as periodic, with periodic boundary conditions applied. As COMSOL adopts the $e^{j\omega t}$ convention, the obtained scattering parameters were post-processed by multiplication with $e^{jk_0 H_{\rm d}}$, where $H_{\rm d} \approx 5\lambda_0$ is the height of the simulation box. This de-embeds the results and repositions the reference ports at the nanodisk's middle plane. For the eigenvalue problem, a COMSOL eigensolver was employed alongside scattering boundary conditions, replacing the ports at the top and bottom of the periodic box.

\begin{table}[h]
\centering
\resizebox{0.6\textwidth}{!}{%
\begin{tabular}{|c|c|c|c|c|}
\cline{2-5}
 \multicolumn{1}{l|}{}& $\varepsilon_{\rm r}$ & $\varepsilon_{\rm i}$ & $\varepsilon_{\rm a}'$ & $\varepsilon_{\rm a}''$ \\ \hline
\textbf{Proof of Concept}          & 8     & 0.01           & 0.1    & 0     \\ \hline
\textbf{Bi3YIG}\cite{Meta-BiYIG-material} ($\lambda=650$ nm) & 8.077 & 0.016          & 0.0733 & 0.007 \\ \hline
\textbf{YIG}\cite{Yig_properties} ($\lambda=1267$ nm)   & 4.84  & $6.6\cdot 10^{-7}$ & 0.0004 & 0     \\ \hline
\end{tabular}}
\caption{Optical parameters used in the study.}
\label{table:Optical_params}
\end{table}

For both BIG and YIG nanodisk simulations, the magnetisation data obtained from MuMax micromagnetic simulations were imported into COMSOL and incorporated into the permittivity tensor~(\ref{eq: Permittivity tensor comsol}), using linear interpolation. The values of the permittivity tensor components were given in Table~\ref{table:Optical_params}. Where $\varepsilon_a$ has been decomposed into the form $\varepsilon_{\rm a}=\varepsilon_{\rm a}^{'}-j\varepsilon_{\rm a}^{''}$.
In the BIG nanodisk case, the smaller size of the nanodisks results in distinct magnetisation profiles for disks with diameters $l_{\rm a}$ and $l_{\rm b}$. Therefore, a separate magnetisation field was used for each nanodisk size. In contrast, for the YIG implementation shown in Supplementary Section~9, the nanodisks have a much larger diameter ($D \simeq 600$ nm), and the difference in diameter between disk types is minimal ($\Delta \simeq 10$ nm). This ensures that the magnetisation fields are nearly identical, allowing the same magnetisation field to be used for both nanodisk types. In both implementations, adjacent nanodisks were assigned opposite vortex polarities (but same circularities), in agreement with the micromagnetic simulation results in Fig.~5a.
\subsection*{Micromagnetic simulations}
Micromagnetic simulations were carried out using Mumax3 (version 3.10), an open-source, GPU-accelerated software tool \cite{Vansteenkiste2014}. 
The input material parameters of BIG, YIG, and cobalt (at room temperature) used in the simulations are provided in Table~\ref{tab-param}. The cubic magnetocrystalline easy axes for both YIG and BIG are oriented along the $[111]$ direction. The exchange length is determined using the expression $l_{\rm ex}=\sqrt{2A_{\rm ex}/\mu_0 M_{\rm s}^2}$. For the cobalt, the hexagonal close-packed (hcp) phase is considered, assuming a uniaxial magnetocrystalline anisotropy with the easy axis oriented along the [100] direction.
\begin{table}[h]
    \centering
    \renewcommand{\arraystretch}{1.5}
    \begin{tabular}{l c c c c c c}
     \hline\hline
        & \multicolumn{1}{c}{\textbf{$M_{\rm s}$}} 
        & \multicolumn{1}{c}{\textbf{$A_{\rm ex}$}} 
        & \multicolumn{1}{c}{\textbf{$K_{\rm c1}$}}
        & \multicolumn{1}{c}{\textbf{$K_{\rm u}$}}
        & \multicolumn{1}{c}{\textbf{$\alpha$}}
        & \multicolumn{1}{c}{\textbf{$l_{\rm ex}$}}\\ 
         \hline
        \textbf{YIG \cite{losby2015torque,prabhakar2009spin}} & $140 \, \text{kA/m}$ & 
        $2 \,\text{pJ/m}$ & $-610 \, \mathrm{J/m^3}$ & $0$ &$0.008$ & $12.7\, \mathrm{nm}$ \\ 
         \hline
         \textbf{BIG \cite{PhysRevB.106.134401}} & $135 \, \text{kA/m}$ & 
        $3.5 \,\text{pJ/m}$ & $-400 \, \mathrm{J/m^3}$  & $-5750 \, \mathrm{J/m^3}$ & 0.005 & $17.5 \, \mathrm{nm}$ \\ 
        \hline
        \textbf{Cobalt \cite{coey_magnetism_nodate,wen-bing_micromagnetic_2010}} & $1.4 \, \text{MA/m}$ & 
        $31 \,\text{pJ/m}$ & $0 $  & $410 \, \mathrm{KJ/m^3}$ & $0.015$ & $5.02 \, \mathrm{nm}$ \\ 
         \hline\hline
    \end{tabular}
        \caption{Material parameters for YIG, BIG and cobalt.}\label{tab-param}
\end{table}

In all simulations, a discretised cell size of $10\, \mathrm{nm} \times 10\, \mathrm{nm} \times 10\, \mathrm{nm}$ was used. This corresponds to a cell volume of approximately $0.49\, l_{\mathrm{\rm ex}}^3$ for YIG and $0.19\, l_{\mathrm{\rm ex}}^3$ for BIG.

\section*{Acknowledgement}
IF, RC and AD acknowledge the Swedish Research Council for Sustainable Development (Formas) (Project No. 2021-01390), Thuréus Forskarhem och Naturminne Foundation, and the Swedish Research Council (VR) (Project No. 2024-05025); BA and VA acknowledge Research Council of Finland grant no. 356797 and Research Council of Finland Flagship Program (Grant No.: 320167, PREIN); VA acknowledges Finnish Foundation for Technology Promotion. ADR and LMME acknowledge the Spanish National Research Council (grant No. PID2021-128442NA-I00 and CNS2024-154715). LMME acknowledges Universitat Politècnica de València (PAID-01-23)

\printbibliography

\end{document}